# Effects of Multi-Surface Modification on Curie temperature of ferroelectric films

Xiao-Guang Wang*
Laboratory of optical physics, Institute of Physics, Chinese Academy of Sciences, Beijing 100080, P.R.China

Shao-Hua Pan and Guo-Zhen Yang
China Center of Advanced Science and Technology (World Laboratory), P.O.Box 8730, Beijing 100080,P. R. China and Institute of Physics, Chinese Academy of Sciences, Beijing 100080, P.R. China

**Abstract**. Within the framework of mean field theory, we study the effects of multi-surface modification on Curie temperature of ferroelectric films using the transverse Ising model. The general nonlinear equations for Curie temperature of multi-surface ferroelectric films with arbitrary exchange constants and transverse fields are derived by the transfer matrix method. As an example, we consider a $(l,n,l)$ film consisting of $l$ top surface layers, $n$ bulk layers and $l$ bottom surface layers. Two types of surface modifications, modifications of a surface exchange constant and a surface transverse field are taken into account. The dependence of Curie temperature on the surface layer numbers, bulk layer numbers, surface exchange constants ,surface transverse fields and bulk transverse fields is discussed.

## 1 Introduction

Size effects are expected to be particular important in materials that undergo phase transitions, such as ferromagnets, superconductors and ferroelectrics[1-3]. In recent years, there has been considerable interest in the effects of size and surface on the ferroelectric phase transition due to the developments of ferroelectric films and composite materials [4-20]. Nonvolatile memories and other applications have made ferroelectric films a subject of great practical importance [21].

An early study on KDP fine particles embedded in an insulating medium showed no ferroelectric phase transition if their size is less than 150 nm [22], while Anliker et al. demonstrated that the transition temperature of $BaTiO_3$ increases as the particle size decreases [23]. Controversial results obtained from $BaTiO_3$[24] fine particles and $PbTiO_3$ fine particles[25] show that the transition temperature decrease with decrease in grain size.

Theoretically, much work has been done on the Laudau phenomenological theory of phase transition in ferroelectric films [6,8,10,11]. On the microscopic level, the pseudo-spin theory based on the transverse Ising model (TIM) was first introduced by de Gennes [26] to describe the phase transition of hydrogen-bonded ferroelectrics. The model has been applied to many other systems such as ferromagnetic systems [27] and Jahn-teller systems [28]. Surface exchange constant and surface layer thickness dependence of Curie temperature of ultra-thin ferroelectric films is studied numerically by considering two types of surface modification of the exchange constants [29]. Curie temperature for ferroelectric films of arbitrary thickness has been considered and modification of both surface exchange constants and surface transverse field is included [30,31]. The effects of few surface layers ($l = 1,2$) of ferroelectric $(l,n,l)$ films on Curie temperature are studied in previous works. Little theoretical work has been devoted to the effects of multi-surface modification ($l \geq 3$) on the Curie temperature. In practice, there exist multiple surface layers in ferroelectric films fabricated by present experimental techniques.

*Email:xyw@aphy.iphy.ac.cn

In this article, we study the ferroelectric $(l,n,l)$ films of arbitrary surface layers $l$ and bulk layers $n$ and take into account two types of surface modification, modification of surface exchange constant $J_s$ and surface transverse field $\Omega_s$. By using the transfer matrix method introduced for ferromagnetic superlattices[32], explicit and general nonlinear equation for Curie temperature is obtained. We give the equation in the following section. The dependence of Curie temperature on surface layer numbers, bulk layer numbers, surface exchange constants and surface transverse fields are discussed in section 3 and the conclusion is given in the last section.

## 2. Transverse Ising model and transfer matrix method

We start with the TIM[26,29,30]

$$H = -\frac{1}{2}\sum_{(i,j)}\sum_{(r,r')} J_{ij} S_{ir}^z S_{jr'}^z - \sum_{ir}\Omega_i S_{ir}^x, \qquad (1)$$

where $S_{ir}^x, S_{ir}^z$ are the $x$ and $z$ components of the pseudo-spin, $(i,j)$ are plane indices and $(r,r')$ are different sites of the planes, $J_{ij}$ denote the exchange constants. We assume that the transverse field $\Omega_i$ is dependent only on layer index and consider the interaction between neighboring sites.

The spin average $\langle \vec{S}_i \rangle$, obtained from the mean field theory [29,30]

$$\langle \vec{S}_i \rangle = \frac{\vec{H}_i}{2|\vec{H}_i|}\tanh(|\vec{H}_i|/2k_B T), \qquad (2)$$

where $\vec{H}_i(\Omega_i, 0, \sum_j J_{ij}\langle S_j^z \rangle)$ is the mean field acting on the ith spin, , $k_B$ is the Boltzman constant and $T$ is the temperature.

At a temperature close and below the Curie temperature, $\langle S_i^x \rangle$ and $\langle S_i^z \rangle$ are small, $|\vec{H}_i| \approx \Omega_i$, equation (2) can be approximated as

$$\langle S_i^x \rangle = (1/2)\tanh(\Omega_i/2k_B T) \qquad (3)$$

$$\langle S_i^z \rangle = \left[z_0 J_{ii}\langle S_i^z \rangle + z(J_{i,i+1}\langle S_{i+1}^z \rangle + J_{i,i-1}\langle S_{i-1}^z \rangle)\right](1/2\Omega_i)\tanh(\Omega_i/2k_B T). \qquad (4)$$

Here $z_0$ and $z$ are the numbers of nearest neighbors in a certain plane and between successive planes respectively. By defining

$$K_i = z_0 J_{ii}/zJ, K_{i,i+1} = J_{i,i+1}/J, K_{i,i-1} = J_{i,i-1}/J, m_i = \langle S_i^z \rangle$$

$$\tau_i = (2\Omega_i/zJ)\coth(\Omega_i/2k_B T), \qquad (5)$$

equation(4) can be written as

$$(\tau_i - K_i)m_i - K_{i,i+1}m_{i+1} - K_{i,i-1}m_{i-1} = 0. \qquad (6)$$

Here $J$ is the interlayer exchange constant in the bulk.

Let us rewrite the above equation in matrix form in analogy with Ref.[32]

$$\begin{pmatrix} m_{i+1} \\ m_i \end{pmatrix} = M_i \begin{pmatrix} m_i \\ m_{i-1} \end{pmatrix} \qquad (7)$$

with $M_i$ as the transfer matrix defined by

$$M_i = \begin{pmatrix} (\tau_i - K_i)/K_{i,i+1} & -K_{i,i-1}/K_{i,i+1} \\ 1 & 0 \end{pmatrix}. \quad (8)$$

We assume that the ferroelectric film contains $N+1$ layers with layer indices $i = 0,1,2...N$. From equation (7), we get

$$\begin{pmatrix} m_N \\ m_{N-1} \end{pmatrix} = R \begin{pmatrix} m_1 \\ m_0 \end{pmatrix}, \quad (9)$$

where $R = M_{N-1}......M_2 M_1$ represents successive multiplication of the transfer matrices.

For an ideal film system, there exists symmetry in the direction perpendicular to the surface, which allows us to write $m_i = m_{N-i}$. Then, the following nonlinear equations for Curie temperature can be obtained from equations (6) and (9) as

$$R_{11}[(\tau_0 - K_0)/K_{0,1}]^2 + (R_{12} - R_{21})[(\tau_0 - K_0)/K_{0,1}] - R_{22} = 0. \quad (10)$$

The above equation is the general equation for Curie temperature for arbitrary exchange constants $J_{ij}$ and transverse field $\Omega_i$.

For a uniform system of $N+1$ layers with $J_{ij} = J$ and $\Omega_i = \Omega$, equation (10) reduces to

$$R_{11}(\tau - z_0/z)^2 + (R_{12} - R_{21})(\tau - z_0/z) - R_{22} = 0, \quad (11)$$

where $\tau = (2\Omega/zJ)\coth(\Omega/2k_B T)$. The total transfer matrix

$$R = \begin{pmatrix} \tau - z_0/z & -1 \\ 1 & 0 \end{pmatrix}^{N-1} \quad (12)$$

The Nth power of a unimodular $2\times 2$ matrix $A$ can be simplified as [28]

$$A^N = U_N A - U_{N-1} I, \quad (13)$$

where $I$ is the $2\times 2$ unit matrix and $U_N = (\lambda_+^N - \lambda_-^N)/(\lambda_+ - \lambda_-)$. $\lambda_+$ and $\lambda_-$ are the two eigenvalues of the matrix $A$.

From equations (12) and (13), we obtain the matrix elements of $R$ as follows

$$\begin{aligned} R_{11} &= (\tau - z_0/z)U_{N-1} - U_{N-2}, \\ R_{12} &= -U_{N-1}, \ R_{21} = U_{N-1}, \\ R_{22} &= -U_{N-2}. \end{aligned} \quad (14)$$

Here, $\lambda_{\pm} = [\tau - z_0/z \pm \sqrt{(\tau - z_0/z)^2 - 4}]/2$. From the definition of $U_N$ and relations $\lambda_+ + \lambda_- = \tau - z_0/z$ and $\lambda_+ \lambda_- = 1$, we have the recursion relations

$$U_N = (\lambda_+ + \lambda_-)U_{N-1} - U_{N-2}, \quad (15)$$

$$U_{N+2} = (\lambda_+ + \lambda_-)^2 U_N - 2(\lambda_+ + \lambda_-)U_{N-1} + U_{N-2} \quad (16)$$

Substituting equation (14) to equation (11) and using the above relations, we reduce the equation (11) to its simplest form

$$U_{N+2} = 0 \quad (17)$$

which is identical with the result of reference [26].

$U_{N+2}$ can be written as

$$U_{N+2} = \sin[(N+2)\phi]/\sin\phi \quad (18)$$

for $(\tau - z_0/z)^2 \leq 4$. Here $\phi = \arccos[(\tau - z_0/z)/2]$. For $(\tau - z_0/z)^2 \geq 4$, $\phi$ becomes $i\theta$, and the trigonometric functions become hyperbolic functions of $\theta$.

Equation (17) gives
$$(2\Omega/zJ)\coth(\Omega/2k_BT) = z_0/z + 2\cos[\pi/(N+2)]. \tag{19}$$
In the limit $N \to \infty$ the bulk Curie temperature $T_B$ is determined by
$$2\Omega\coth(\Omega/2k_BT_B) = n_0 J, \tag{20}$$
where $n_0 = z_0 + 2z$ is the total number of neighbours. It can be seen that equation (20) has a solution only if $n_0 J \geq 2\Omega$.

## 3. Curie temperature of ferroelectric films with multi-surface modification

As an example, we consider a $(l,n,l)$ film consisting of $l$ top surface layers, $n$ bulk layers and $l$ bottom surface layers and study three models. In model I, we assume that pseudo-spins lie on a simple cubic lattice and that the transverse field $\Omega_i = \Omega$. The coupling strength in a surface layer is denoted by $J_s$ and that in a bulk layers or between successive layers is denoted by $J$. Then the transfer matrix (equation(8)) reduces to two different types of matrices

$$P = \begin{pmatrix} X & -1 \\ 1 & 0 \end{pmatrix}, \qquad Q = \begin{pmatrix} Y & -1 \\ 1 & 0 \end{pmatrix} \tag{21}$$

and the total transfer matrix

$$R = P^{l-1} Q^n P^{l-1}. \tag{22}$$

Here $X = \tau - 4J_s/J$ and $Y = \tau - 4$. The nonlinear equation (10) for Curie temperature thus reduces to

$$R_{11}X^2 + (R_{12} - R_{21})X - R_{22} = 0. \tag{23}$$

The matrix elements of $R$ can be obtained explicitly from equations(13) as
$$\begin{aligned}
R_{11} &= V_{l-1}^2 W_n X^2 Y - V_{l-1}^2 W_{n-1} X^2 - 2V_{l-1}V_{l-2}W_n XY + \\
&\quad 2(V_{l-1}V_{l-2}W_{n-1} - V_{l-1}^2 W_n)X + V_{l-2}^2 W_n Y + V_{l-1}^2 W_{n-1} + 2V_{l-1}V_{l-2}W_n - V_{l-2}^2 W_{n-1}, \\
R_{12} &= -R_{21} = -V_{l-1}^2 W_n XY + (V_{l-1}^2 W_{n-1} + V_{l-1}V_{l-2}W_n)X + V_{l-1}V_{l-2}W_n Y \\
&\quad + V_{l-1}^2 W_n - 2V_{l-1}V_{l-2}W_{n-1} - V_{l-2}^2 W_n \\
R_{22} &= -V_{l-1}^2 W_n Y + V_{l-1}^2 W_{n-1} + 2V_{l-1}V_{l-2}W_n - V_{l-2}^2 W_{n-1}.
\end{aligned} \tag{24}$$

Here $V_l = (\alpha_+^l - \alpha_-^l)/(\alpha_+ - \alpha_-)$, $W_n = (\beta_+^n - \beta_-^n)/(\beta_+ - \beta_-)$, $\alpha_\pm = (X \pm \sqrt{X^2 - 4})/2$ and $\beta_\pm = (Y \pm \sqrt{Y^2 - 4})/2$.

For the special case of $l = 1$, equation (20) reduces to
$$W_{n+1}X^2 - 2W_n X + W_{n-1} = 0 \tag{25}$$
which is the same as that of Ref.[30] for the structure with one top and one bottom surface layer. In obtaining equation (25), we have used equation (15).

Equation (23) can be numerically solved for arbitrary $l$ and $n$. Figure 1 shows the reduced Curie temperature $t_C = k_B T_C/J$ as a function of $J_s/J$ for films $(l,n,l)$ with fixed bulk layer numbers $n = 1$ but different surface layer numbers $l$. Curie temperature increases with increasing $l$ for larger $J_s/J$, while it is almost independent of $l$ when $J_s/J$ is sufficiently small. Figure 2 shows $t_C$ vs. $J_s/J$ for films $(3,n,3)$ with different bulk layer numbers. The four curves $n$ merge into one curve for larger $J_s/J$, i.e., the Curie temperature are insensitive to the bulk layer numbers $n$ in this case. On the contrary, Curie temperature depends strongly on number $n$ for smaller $J_s/J$ (say $J_s/J < 1$) and $t_C$ increases as $n$ increases.

Only one type of surface modification, the modification of surface exchange constant $J_s$ is taken into account in the above discussion. Next we study model II including the modification of surface transverse field. Let $\Omega_s$ and $\Omega$ denote transverse field in surface and bulk layers respectively. The nonlinear equations for Curie temperature is still the equations (23) and (24) except for the change of the element $X$ from $\tau - 4J_s/J$ to $\tau_s - 4J_s/J$, where $\tau_s = (2\Omega_s/J)\coth(\Omega_s/2k_B T)$. In figure 3, we give the Curie temperature via $\Omega_s/J$. The Curie temperature decreases as $\Omega_s/J$ increases. The transverse field causes a reduction in the Curie temperatures of the film.

Now we consider a more realistic model, model III, in which we assume $J_{ij} = J_s$ for both sites in surface layers, $J_{ij} = J$ for both sites in bulk layers, and $J_{ij} = J_{BS} = \sqrt{J_s J}$ between interface layers. The transverse fields $\Omega_i = \Omega_s$ for sites in the surface layers, $\Omega_i = \Omega$ for sites in the bulk layers, and $\Omega_i = \Omega_{BS} = \sqrt{\Omega_s \Omega}$ in the interface layers. The total transfer matrix model III becomes

$$R = M_S^{l-2} M_{SB} M_B^{n-2} M_{BS} M_S^{l-2}, \qquad (26)$$

where

$$M_S = \begin{pmatrix} \tau_s J/J_s - 1 & -1 \\ 1 & 0 \end{pmatrix}$$

$$M_B = \begin{pmatrix} \tau - 1 & -1 \\ 1 & 0 \end{pmatrix}$$

$$M_{BS} = M_l M_{l-1} = \begin{pmatrix} (\tau_{BS}-4)(\tau_{BS}-4J_s/J)J/J_{BS} - J_{BS}/J & -(\tau_{BS}-4)J_s/J_{BS} \\ (\tau_{BS}-4J_s/J)J/J_{BS} & -J_s/J_{BS} \end{pmatrix}$$

$$M_{SB} = M_{l+n} M_{l+n-1} = \begin{pmatrix} (\tau_{BS}-4)(\tau_{BS}-4J_s/J)J^2/(J_s J_{BS}) - J_{BS}/J_S & -(\tau_{BS}-4)J_s/J)J^2/(J_s J_{BS}) \\ (\tau_{BS}-4)J/J_{BS} & -J/J_{BS} \end{pmatrix}$$

(27)

and $\tau_{BS} = 2\Omega_{BS}/J \coth(\Omega_{BS}/2K_B T)$. For film $(2,2,2)$, the transfer matrix reduces to $M_{SB} M_{BS}$. From equations (10) and (27), we can numerically calculate the Curie temperature of $(l,n,l)$ films. In order to compare with the results of model II, we consider the Curie temperature vs. $\Omega_s/J$ and choose parameters the same as those in figure 3. The results are given in figure 4. For small $\Omega_s/J$ (say $\Omega_s/J = 0.1$), the Curie temperatures in model III are larger than those in model II. It is clear that the results are qualitatively consistent with those of model II, while they are quatitatively different. For large $\Omega_s/J$ (say $\Omega_s/J = 2.5$) and $J_s/J = 0.1$, the Curie temperatures of model III are smaller than those in model II. It can be seen from both the figure 3 and figure 4 that the Curie temperature increases as $J_s/J$ increases. Figure 5 gives the Curie temperature vs. surface layer number $l$. The solid line in the figure corresponds to the Curie temperature $t_C^S$ of bulk system with $J_{ij} = J_s$ and $\Omega_{ij} = \Omega_s$. Let $\Omega \to \Omega_s, J \to J_s$ in equation(20), the temperature $t_C^S$ can be calculated. As seen from the figure, the Curie temperature increases with the increase of surface layer number and approach asymptotically to the bulk temperature $t_C^S$.

## 4. Conclusions

By making use of the transfer matrix method, an analytical and general nonlinear equation for Curie temperature of ferroelectric films with arbitrary exchange constant $J_{ij}$ and transverse field $\Omega_i$ has been derived based on the TIM. The dependence of Curie temperature on surface layer numbers, bulk layer numbers, surface exchange constants, surface transverse fields and bulk transverse fields can be easily studied by the nonlinear equation. As an example, we study three models and investigate the effects of multi-surface modification on Curie temperature of symmetric ferroelectric films $(l, n, l)$. In addition, the method proposed here can be applied to phase transition of both infinite and finite ferroelectric superlattices. It has a favorable point because of the simplicity of the numerical calculations.

Figure Captions:

Figure 1 The reduced Curie temperature $t_C$ vs. $J_s/J$ in model I for films $(l,1,l)$ with $l=1$ (solid line); $l=2$ (dotted line), $l=3$ (dashed line), and $l=10$ (dot-dashed line). The parameter $\Omega/J = 2.0$.

Figure 2 The same as figure 1, but for films $(3,n,3)$ with $n=2$ (solid line), $n=5$ (dotted line), $n=10$ (dashed line) and $n=20$ (dot-dashed line). The parameter $\Omega/J = 2.5$.

Figure 3 The reduced Curie temperature vs. $\Omega_s/J$ in model II for films $(3,2,3)$ with $J_s/J$ =0.1(solid line), $J_s/J$ =0.5(dotted line), $J_s/J$ =1.0(dashed line) and $J_s/J$ =1.5(dot-dashed line). The parameter $\Omega/J = 2.5$.

Figure 4 The same as figure 3, but for model III.

Figure 5 The reduced Curie temperature vs. $l$ in model III. The parameters $\Omega_s = 2.0, \Omega = 2.5, J_s = 1.2$ and $n=5$.

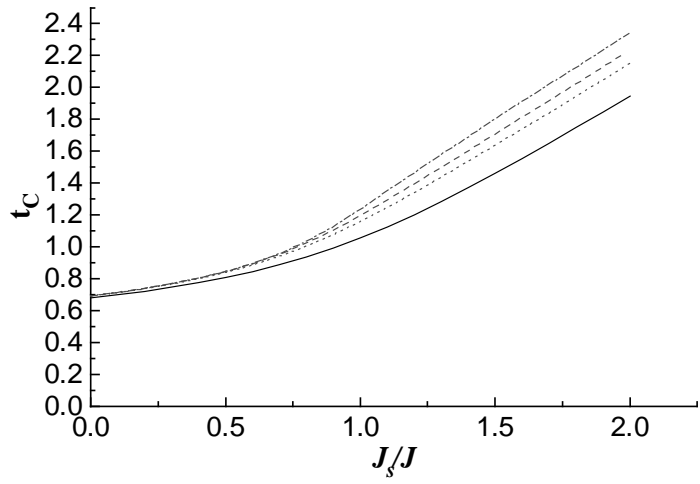

Figure1

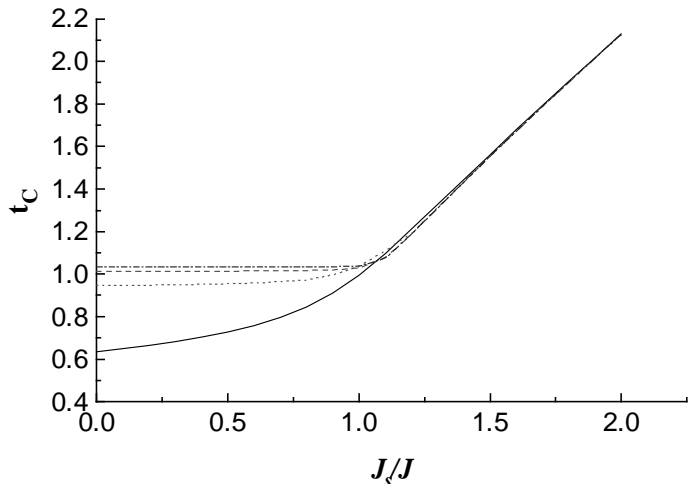

Figure2

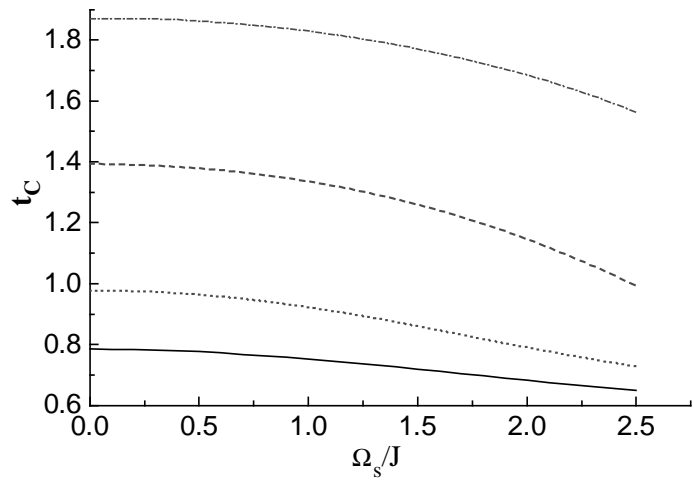

Figure 3

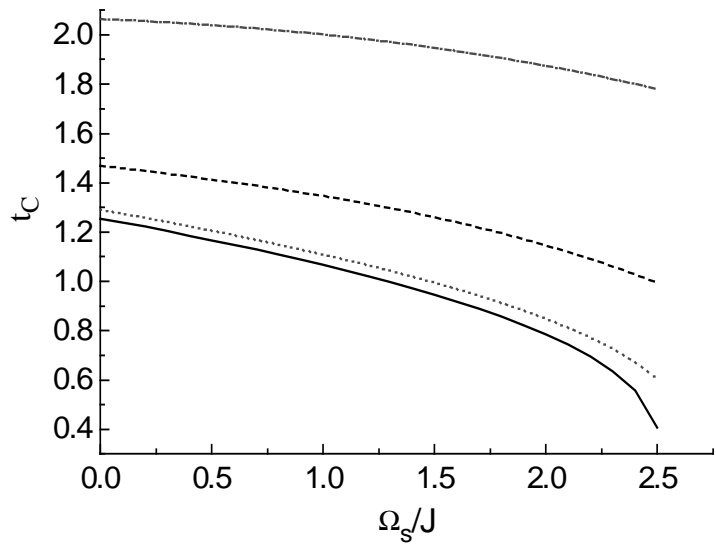

Figure 4

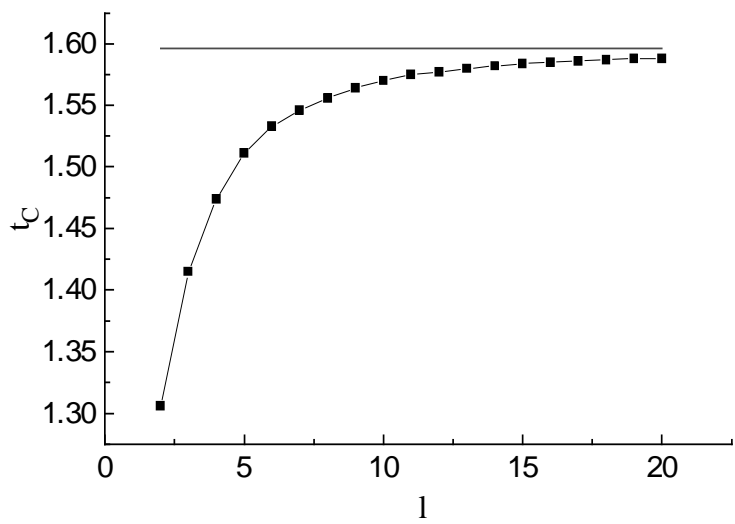

Figure 5